\documentclass{ol-softwaremanual}

\usepackage[utf8]{inputenc}
\usepackage[english]{babel}
\usepackage{multirow}
\usepackage[table,xcdraw]{xcolor}
\usepackage{graphicx}
\usepackage{fancyhdr}
\usepackage{lastpage}
\usepackage{graphicx}
\usepackage{enumitem}
\usepackage{listings}
\usepackage{color}
\usepackage{appendix}
\usepackage{hyperref}
\usepackage[ruled]{algorithm2e}
\usepackage{todonotes}
\usepackage{setspace}
\usepackage{lipsum}
\usepackage{caption}
\usepackage{subcaption}
\usepackage{csquotes}
\usepackage[right]{lineno}
\usepackage{amsmath,amsfonts}
\usepackage{cleveref}
\usepackage{lineno}

\usepackage[
backend=biber,
style=numeric,
sorting=ynt
]{biblatex}
\renewcommand\_{\textunderscore\allowbreak}

\let\oldcenter\center
\let\oldendcenter\endcenter
\renewenvironment{center}{\setlength\topsep{0pt}\oldcenter}{\oldendcenter}

\definecolor{dkgreen}{rgb}{0,0.6,0}
\definecolor{gray}{rgb}{0.5,0.5,0.5}
\definecolor{mauve}{rgb}{0.58,0,0.82}
\definecolor{backcolour}{rgb}{0.95,0.95,0.92}

\usepackage{lipsum}
\newcommand{\subsubsubsection}[1]{\paragraph{#1}\mbox{}\\}
\usepackage{tikz}
\usetikzlibrary{calc,backgrounds,shapes}
\usetikzlibrary{arrows.meta, shapes.callouts}
\usetikzlibrary{decorations.pathreplacing}

\tikzset{global scale/.style={
    scale=#1,
    every node/.style={scale=#1}
  }
}

\let\state\textsf
\newcommand\ESTABLISHED{\state{ESTABLISHED}}
\newcommand\SYNRECEIVED{\state{SYN-RECEIVED}}
\newcommand\CLOSED{\state{CLOSED}}
\newcommand\CLOSEWAIT{\state{CLOSE-WAIT}}
\newcommand\LISTEN{\state{LISTEN}}
\newcommand\SYNSENT{\state{SYN-SENT}}
\newcommand\FINWAITONE{\state{FIN-WAIT-1}}
\newcommand\FINWAITTWO{\state{FIN-WAIT-2}}
\newcommand\CLOSING{\state{CLOSING}}
\newcommand\LASTACK{\state{LAST-ACK}}
\newcommand\TIMEWAIT{\state{TIME-WAIT}}

\usepackage{lipsum}
\title{Security-Hardening Software Libraries with Ada and SPARK \\[0.5em] \LARGE{A TCP Stack Use Case} \\[1em] \Large{White Paper}}
\author{Kyriakos Georgiou, Guillaume Cluzel, Paul Butcher, Yannick Moy \\[1em]}
\softwarelogo{\includegraphics[width=8cm]{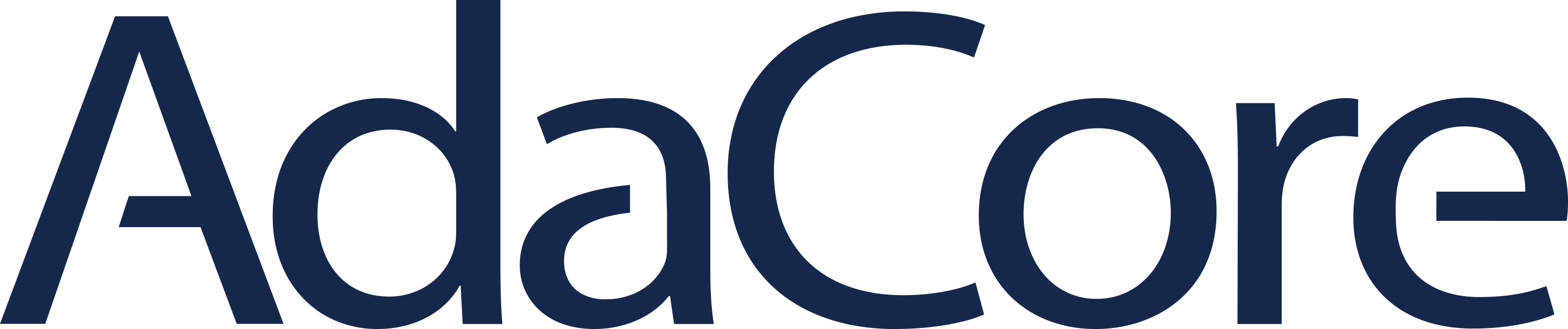}}

\begin{document}
\maketitle

\begin{abstract}
The work is part of a series of white papers to demonstrate how the SPARK technology, a subset of the Ada programming language supported by formal verification tools, can be applied for the security-hardening of Software libraries. The first white paper of this series, \cite{AdaCore_Hardening_Soft_Libr_EMBENCH}, introduced the SPARK technology through the conversion of a \emph{C} benchmark suite to SPARK. The work demonstrated how the use of the SPARK technology could guarantee the absence of runtime errors without significantly impacting performance when compared to the \emph{C} version of the code. Furthermore, best practices and guidelines on achieving the absence of runtime errors with SPARK were reported. This white paper builds on the previous work and demonstrates how an existing professional-grade, open-source embedded TCP/IP stack implementation written in the C programming language can be hardened using the SPARK technology to increase its assurance, reliability, and security. More specifically, the work demonstrates the ability of SPARK to enforce the correct usage of the library and to verify the conformance to its functional specifications. A multifaceted approach achieves this. Firstly, the TCP layer's C code is being replaced with formally verified SPARK. Then the lower layers, still written in C and on which the TCP layer depends, are modeled using SPARK contracts and validated using symbolic execution with KLEE. Finally,  formal contracts for the upper layers are defined to call the TCP layer. The work allowed the detection and correction of two bugs in the TCP layer. The powerful approach detailed in this work can be applied to any existing critical C library.
\end{abstract}

\newpage

\tableofcontents
\newpage

\section{Introduction}

TCP is the most widely used network protocol to communicate on the Internet. Thus, ensuring the TCP/IP stack's safety is an essential step towards safer cyber-physical systems. Existing research deals with formally verifying protocols of other TCP/IP stack levels. For example, the work in \emph{miTLS} \cite{bhargavan2013implementing} formally verifies an SSL/TLS protocol implementation, and the work in \cite{Reiher_2020} uses a technology called RecordFlux to safely parse data segments. At this point of writing, we are not aware of any formally verified TCP protocol implementation. Such an implementation would provide strong assurances on the security and the safety of a TCP implementation. Furthermore, the security of higher-level protocols in the TCP/IP stack,  such as the TLS protocol, can only be ensured if the underlying TCP implementation is bug-free and conforms with its functional specifications.

SPARK 2014 is a programming language designed as a subset of Ada to produce highly reliable software through formal verification. SPARK can detect uninitialized variables with control flow analysis, ensure the absence of run-time errors, and, based on SMT-solver technology, provides mechanisms to specify and mathematically prove the functional behavior of a program.

This work aims to enhance the safety and security of an embedded industrial-grade implementation of the TCP protocol by using the SPARK technology.

To fully grasp the technical parts of this work, we urge the reader to first have a read at the first white paper on hardening software libraries with SPARK \cite{AdaCore_Hardening_Soft_Libr_EMBENCH}. The SPARK syntax, usage, assurance levels, and bottom-up approach of introducing it into a C-based project are described in great detail.

\section{The TCP protocol}

This section is not meant to define the full functional specification of the TCP protocol. Instead, the intention is to give enough context to enable the reader to follow the work conducted in this deliverable. For a full definition of the protocol, please refer to the \emph{RFC 793} document [1], the defining standard for TCP.

\subsection{Networking Communication Models}

Networking communication models define the processes of transferring information from one network component to another. Currently, the two most popular networking communication models are the TCP/IP (Transmission Control Protocol/Internet Protocol) \cite{TCP_IP:ietf_tutorial} and the OSI (Open System Interconnection) \cite{ISO:35.100} models, with the first one being the most widely used. The TCP/IP model was developed by the US Department of Defence (DoD) \cite{DoD} and the OSI model was introduced by ISO (International Standard Organization) \cite{ISO}. Their purpose is to provide hardware vendors with a common base for the development of networking products that can communicate with each other.

\begin{figure}
    \centering
    \resizebox{\textwidth}{!}{%
    \centering
    \includegraphics[trim={0cm 7cm 2.5cm 0},clip]{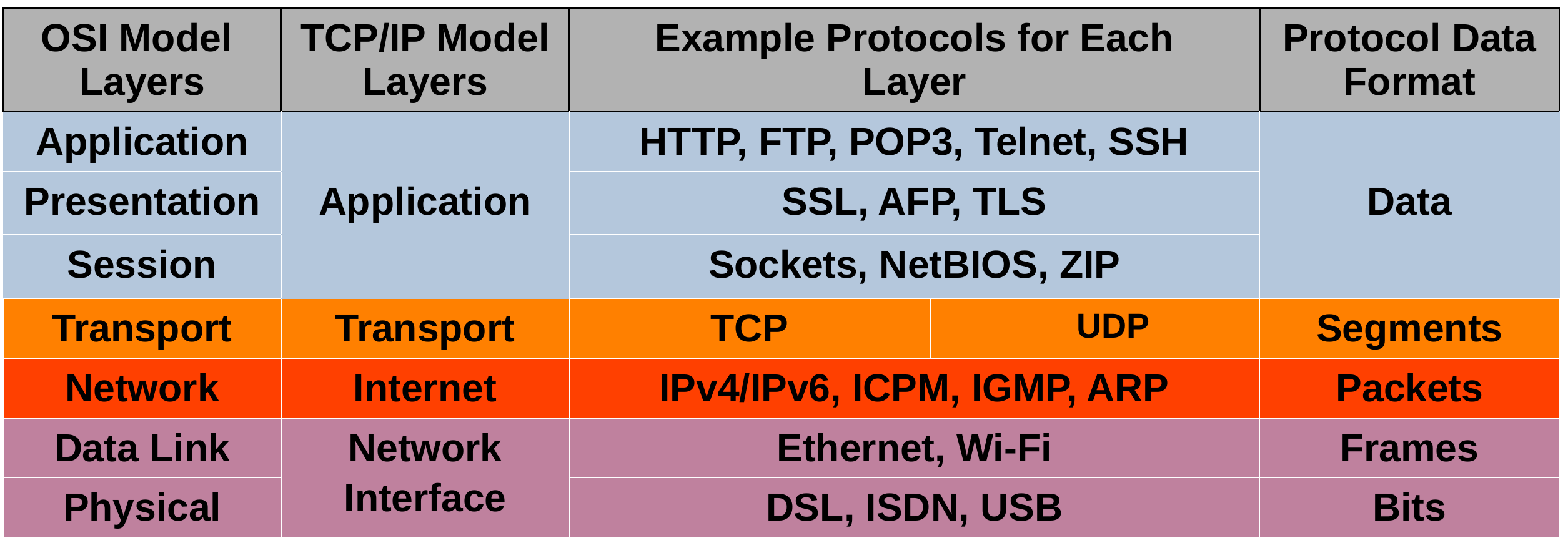}
    }
    \caption{Overview of the TCP/IP and the OSI networking models for communication}
    \label{fig:TCP-IP-structure}
\end{figure}

\Cref{fig:TCP-IP-structure} shows an overview of the two models, TCP/IP and OSI. TCP/IP has four layers, while the OSI is a 7-layer model. Layering is an essential design requirement for both models to achieve modularity, flexibility, and abstraction. Applications that only need to use the functionality of the lower levels can drop all the unnecessary upper levels. Furthermore, layering enables easier troubleshooting as the network's functionality is well-defined and split between the different layers. Each layer supports a wide range of protocols that are compatible with the layer's specifications. The lowest layers, bottom layers in \Cref{fig:TCP-IP-structure}, are more hardware-oriented, while the top layers are more software specific. Software developers producing networking applications have more responsibility for the implementation and usage of the TCP/IP \emph{Transport} and \emph{Application} layers, or their corresponding OSI model's layers. While the \emph{Application} level has a wide range of protocols that can be used depending on the application specification, the \emph{Transport} layer is heavily depended on the two dominant protocols:

\begin{description}[wide = 0pt]
    \item[\state{1. Transmission Control Protocol (TCP)}] is a connection-oriented protocol; a connection between the sender and the receiver must be established before transmitting any data. Furthermore, TCP is a reliable protocol since it guarantees the delivery of all the messages and that they are delivered in the order they were sent. It also provides error-checking mechanisms to discard and recover any corrupted data. 
    \item[\state{2. User Datagram Protocol (UDP)}] is a connectionless protocol; no connection establishment is required between the sender and the receiver. UDP is more performance-oriented rather than reliability oriented; thus, it does not provide any error-checking mechanism, and it does not guarantee the delivery of datagrams sent.
\end{description}

Today, almost any networking application is using either the TCP or the UDP protocols to traffic data between the Network and the Application layers. Thus, the reliability, assurance and security of these protocols are of paramount importance for critical applications. A security vulnerability in the implementation of these protocols can lead to a severe compromise of the security of any application using them~\cite{zero_days_vuln:ARMIS_white_paper}. Since for critical cyber-physical systems reliability is an important factor, the focus of this work will be on the TCP protocol.

\subsection{\emph{CycloneTCP} TCP/IP library}

\emph{CycloneTCP} is a professional-grade embedded TCP/IP library developed by the \emph{Oryx Embedded} company. The implementation is meant to conform with the Request for Comments (RFC) Internet standards, namely the \emph{RFC 793} TCP protocol specifications \cite{rfc793}, provided by the Internet Engineering Task Force (IETF) \cite{ietf}. The library is written in ANSI C, and it supports a large number of 32-bit embedded processors and a large number of Real-time Operating Systems (RTOS). It can also run on bare metal environments. The library offers implementations for a wide range of TCP/IP protocols. These includes the IPv4 and IPv6 of the network layer, the TCP and UDP of the transport layer, and the DHCP and HTTP of the application layer. A quick overview of the library can be found here \cite{CycloneTCP} while an open-source version can be downloaded at \cite{CycloneTCP:download}. The development of this work is based on the \emph{CycloneTCP}'s Transport layer implementation of the TCP and UDP protocols, with a focus on the TCP protocol. While the TCP protocol is meant to conform with the \emph{RFC 793} TCP Specification \cite{rfc793}, no formal guarantees are being provided on the conformance. Thus, this work takes a step further than the original protocol's implementation to provide some assurances on the specification conformance and the overall security hardening of the library. 

\subsection{Overview of the TCP protocol Specification}

This section gives an overview of the TCP functional specification without going into unnecessary depth. A complete description of the protocol's specification can be found at the \emph{RFC 793} document \cite{rfc793}. TCP is a reliable, ordered, and error-checked connection-oriented protocol; connection must be established before the transmission of any data, and the connection must be closed once all the data has been transmitted.

In the \emph{CycloneTCP} implementation, the \emph{socket} data structure is used to retain the status of a TCP connection. \Cref{fig:socket} shows part of its equivalent implementation in the Ada/SPARK code, demonstrating some of the fields that store vital information for the status of a connection. Furthermore, a socket can be manipulated by the user through the TCP's library Application Programming Interface (API) to perform various TCP operations, such as the transmission of data.

\begin{figure}
\begin{lstlisting}[language=Ada]
    type Socket_Struct is record
    S_Descriptor     : Sock_Descriptor;
    S_Type           : Socket_Type;
    S_Protocol       : Socket_Protocol;
    S_Net_Interface  : System.Address;
    S_localIpAddr    : IpAddr;
    S_Local_Port     : Port;
    S_Remote_Ip_Addr : IpAddr;
    S_Remote_Port    : Port;
    S_Timeout        : Systime;
    State            : Tcp_State;
    -- Other fields
end record;
\end{lstlisting}
    \caption{The \emph{socket} Ada implementation.}
    \label{fig:socket}
\end{figure}

 Generally, every TCP communication session will go through the following three phases (if no error occurs):
\begin{enumerate}
    \item Opening the connection
    \item Sending and receiving the data
    \item Closing the connection
\end{enumerate}

\begin{figure}[p]
    \begin{tikzpicture}[>=latex, global scale=.75]

    %
    %
    \tikzstyle{state} = [draw, very thick, fill=white, rectangle, minimum height=3em, minimum width=7em, node distance=8em, font={\sffamily\bfseries}]
    \tikzstyle{stateEdgePortion} = [black,thick];
    \tikzstyle{stateEdge} = [stateEdgePortion,->];
    \tikzstyle{edgeLabel} = [pos=0.5, text centered, font={\sffamily\small}];

    %
    %
    \node[state, name=closedStart] {CLOSED};
    \node[state, name=listen, below of=closedStart] {LISTEN};
    \node[state, name=synSent, below of=listen, right of=listen, xshift=8em] {SYN\_SENT};
    \node[state, name=synRcvd, below of=listen, left of=listen, xshift=-8em] {SYN\_RECEIVED};
    \node[state, name=established, below of=listen, node distance=14em] {ESTABLISHED};
    \node[state, name=finWait1, below of=established, left of=established, node distance=7em, xshift=-9em] {FIN\_WAIT\_1};
    \node[state, name=finWait2, below of=finWait1] {FIN\_WAIT\_2};
    \node[state, name=closeWait, below of=established, right of=established, node distance=7em, xshift=9em] {CLOSE\_WAIT};
    \node[state, name=closing, below of=established, node distance=14em] {CLOSING};
    \node[state, name=lastAck, below of=closeWait] {LAST\_ACK};
    \node[state, name=timeWait, below of=closing] {TIME\_WAIT};


    %
    %
    \draw ($(closedStart.south) + (-.5em,0)$)
    edge[stateEdge] node[edgeLabel, xshift=-3em]{\emph{Passive open}}
    ($(listen.north) + (-.5em,0)$);
    \draw ($(listen.north) + (.5em,0)$)
    edge[stateEdge] node[edgeLabel, xshift=2em]{\emph{Close}}
    ($(closedStart.south) + (.5em,0)$);

    \draw ($(listen.south) + (-1em,0)$)
    edge[stateEdge, bend left=22.5] node[edgeLabel, xshift=-3em, yshift=1.5em, text width=10em]{rcv/SYN,\\ snd/SYN \& ACK}
    ($(synRcvd.east) + (0,1em)$);
    \draw ($(listen.south) + (1em,0)$)
    edge[stateEdge, bend right=22.5] node[edgeLabel, xshift=1.5em, yshift=1.5em, text width=5em]{\emph{Send},\\ snd/SYN}
    ($(synSent.west) + (0,1em)$);

    \draw ($(synRcvd.north) + (.5em,0)$)
    edge[stateEdge, bend left=45] node[edgeLabel,xshift=-2.5em, text width=6em]{\emph{Timeout},\\ snd/RST}
    ($(closedStart.west) + (0,-.5em)$);

    \draw ($(synSent.north) + (-.5em,0)$)
    edge[stateEdge, bend right=45] node[edgeLabel,xshift=-1em, yshift=-1em]{\emph{Close}}
    ($(closedStart.east) + (0,-.5em)$);
    \draw ($(closedStart.east) + (0,.5em)$)
    edge[stateEdge, bend left=45] node[edgeLabel,xshift=3.5em, text width=8em]{\emph{Active open},\\ snd/SYN}
    ($(synSent.north) + (.5em,0)$);

    \draw (synSent.west)
    edge[stateEdge] node[edgeLabel, yshift=1em]{rcv/SYN, snd/ACK}
    (synRcvd.east);
    \draw (synRcvd)
    edge[stateEdge] node[edgeLabel, xshift=-2em, text width=5em]{\emph{Close},\\ snd/FIN}
    (finWait1);

    \draw ($(synRcvd.east) + (0,-1em)$)
    edge[stateEdge, bend left=22.5] node[edgeLabel, xshift=-1em, yshift=-1em]{rcv/ACK}
    ($(established.north) + (-1em,0)$);
    \draw ($(synSent.west) + (0,-1em)$)
    edge[stateEdge, bend right=22.5] node[edgeLabel, xshift=2.8em, yshift=-1.5em, text width=8em]{rcv/SYN \& ACK,\\ snd/ACK}
    ($(established.north) + (1em,0)$);

    \draw ($(established.south) + (-1em,0)$)
    edge[stateEdge, bend left=22.5] node[edgeLabel, xshift=-1em, yshift=1em, text width=4em]{\emph{Close},\\ snd/FIN}
    ($(finWait1.east) + (0,.5em)$);
    \draw ($(established.south) + (1em,0)$)
    edge[stateEdge, bend right=22.5] node[edgeLabel, xshift=2em, yshift=1em,text width=5em]{rcv/FIN,\\ snd/ACK}
    ($(closeWait.west) + (0,1em)$);

    \draw (finWait1.south)
    edge[stateEdge] node[edgeLabel, xshift=-2em]{rcv/ACK}
    (finWait2.north);
    \draw ($(finWait1.east) + (0,-.5em)$)
    edge[stateEdge, bend left=22.5] node[edgeLabel, xshift=4em, text width=6em]{rcv/FIN,\\ snd/ACK}
    (closing.north);
    \draw (finWait1.south east)
    edge[stateEdge, bend right=10] node[edgeLabel, xshift=1.7em, yshift=2.3em, text width=8em]{rcv/FIN \& ACK,\\ snd/ACK}
    (timeWait.north west);

    \draw (finWait2.south)
    edge[stateEdge, bend right=22.5] node[edgeLabel, xshift=-2em, yshift=-1em, text width=5em]{rcv/FIN,\\ snd/ACK}
    (timeWait.west);

    \draw (closing)
    edge[stateEdge] node[edgeLabel, xshift=2em]{rcv/ACK}
    (timeWait);

    \draw (closeWait)
    edge[stateEdge] node[edgeLabel,xshift=2.5em, text width=6em]{\emph{Close},\\ snd/FIN}
    (lastAck);

    %
    %
    \coordinate (lastAck2ClosedA) at ($(lastAck.east) + (2em,0)$);
    \coordinate (lastAck2ClosedB) at ($(closedStart.north -| lastAck.east) + (2em,1em)$);
    \coordinate (lastAck2ClosedC) at ($(closedStart.north) + (0.5em,1em)$);
    \draw (lastAck.east) edge[stateEdgePortion] (lastAck2ClosedA);
    \draw (lastAck2ClosedA) edge[stateEdgePortion] node[edgeLabel,xshift=-2em, yshift=-4em]{rcv/ACK} (lastAck2ClosedB);
    \draw (lastAck2ClosedB) edge[stateEdgePortion] (lastAck2ClosedC);
    \draw (lastAck2ClosedC) edge[stateEdge] ($(closedStart.north) + (0.5em,0)$);

    %
    %
    \coordinate (timeWait2ClosedA) at ($(timeWait.south) + (0,-1em)$);
    \coordinate (timeWait2ClosedB) at ($(timeWait.south -| finWait2.west) + (-2em,-1em)$);
    \coordinate (timeWait2ClosedC) at ($(closedStart.north -| finWait2.west) + (-2em,1em)$);
    \coordinate (timeWait2ClosedD) at ($(closedStart.north) + (-0.5em,1em)$);
    \draw (timeWait.south) edge[stateEdgePortion] (timeWait2ClosedA);
    \draw (timeWait2ClosedA) edge[stateEdgePortion] (timeWait2ClosedB);
    \draw (timeWait2ClosedB) edge[stateEdgePortion] (timeWait2ClosedC);
    \draw (timeWait2ClosedC) edge[stateEdgePortion]
    node[edgeLabel, text width=12.25em, yshift=1.5em]{\emph{Timeout after two maximum segment lifetimes (2*MSL)}}
    (timeWait2ClosedD);
    \draw (timeWait2ClosedD) edge[stateEdge] ($(closedStart.north) + (-0.5em,0)$);

    \begin{pgfonlayer}{background}
        \draw [join=round,black,dotted] ($(closeWait.north west) + (-1em, -1em)$) rectangle ($(lastAck.south east) + (1em, 1em)$);
        \draw [join=round,black,dotted] ($(finWait1.north west) + (-1em, -1em)$) rectangle ($(timeWait.south east) + (1em, 1em)$);
    \end{pgfonlayer}

    \end{tikzpicture}
    \caption{The TCP connection automaton demonstrates the different states a TCP session can exhibit. The figure is recreated from \cite{TCPautomaton:source}.}
    \label{fig:TCPAutomaton}
\end{figure}
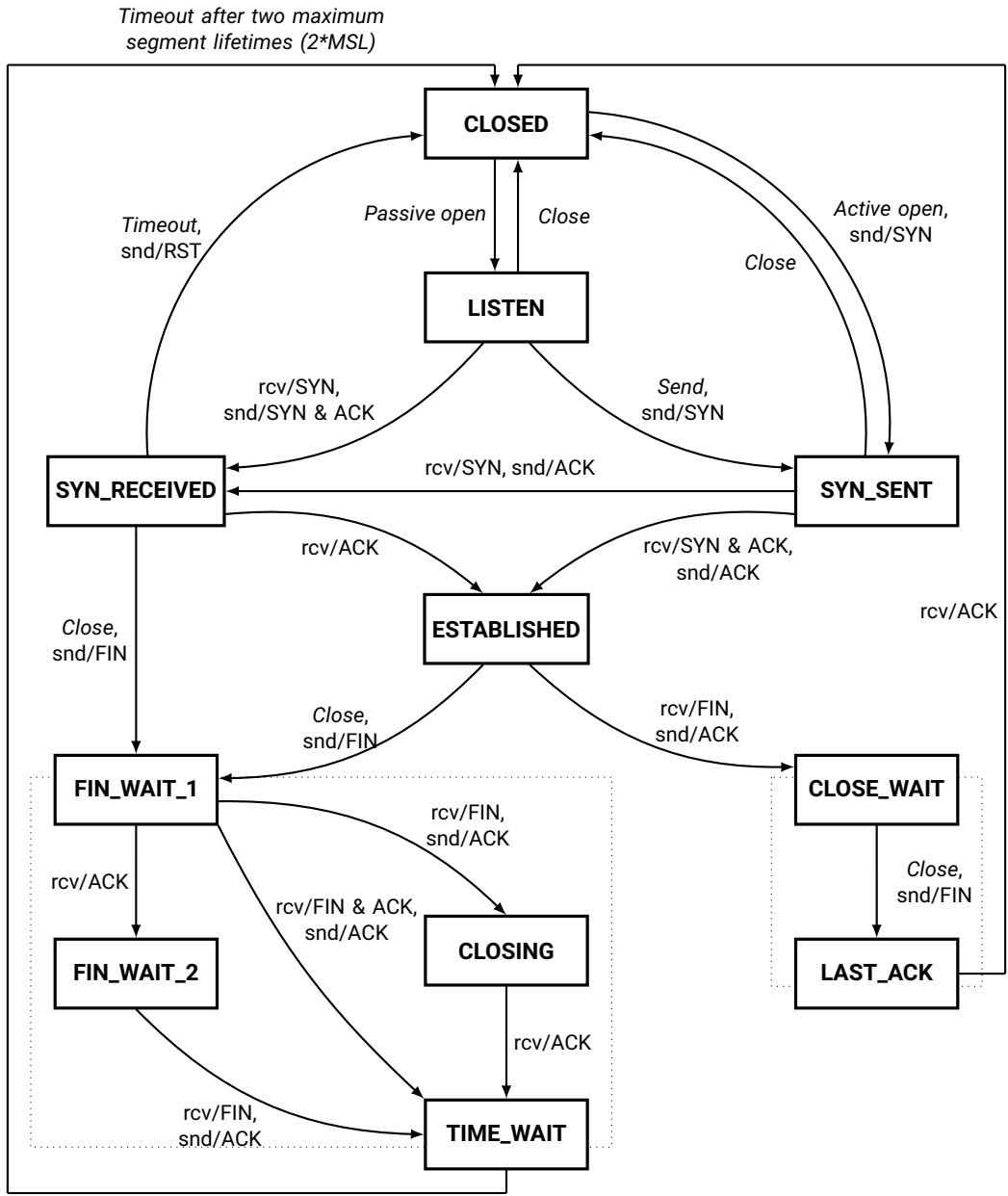

The behaviour of the three-phase communication can be described by the state machine given in \Cref{fig:TCPAutomaton}. An edge represents a state transition. A transition is either triggered by an explicit action, stated in \emph{italics} format on the label's edge, for example, \emph{Close}, or is triggered by the arrival of a specific flag/s. An edge-label in the format of $A/B$ represents the associated flags transmitted with each transition, where $A$ is either send or receive and $B$ refers to the actual flag/s transmitted in response to the action that caused the state-transition. An explicit action is either a user-triggered action (through the library's API) or an automatically performed action triggered by a timeout event. The flags are embedded in the header section of a transmitted segment and can be one of the following:

\begin{itemize}
    \item \textbf{ACK} -- Acknowledgement field significant. The last message received by the sender is acknowledged. 
    \item \textbf{SYN} -- Synchronize sequence number. This flag is sent to establish a connection.
    \item \textbf{FIN} -- No more data from sender. This flag is sent to close the connection.
\end{itemize}

The state machine in \Cref{fig:TCPAutomaton} does not represent the complete protocol specification; for example, it does not reflect error-conditions or any actions which are not connected with the state changes. Rather, it gives an overview of all the possible states a TCP connection could reach over its lifetime. The \emph{socket} data structure given in \Cref{fig:socket} can represent these states.

Finally, although the state machine of \Cref{fig:TCPAutomaton} is a recreation of the TCP state machine found in the TCP specification document on page 22 \cite[23]{rfc793} the two are not entirely identical. The state machine of \Cref{fig:TCPAutomaton} is updated to include an extra transition between the states \FINWAITONE{} and \TIMEWAIT{}. The extra transition reflects the case where the ACK and FIN flags are received in the same segment. It is an allowed behaviour by the TCP specification, which the \emph{CycloneTCP} library also exhibits.

The meaning of the states, as taken from the TCP specification document \cite{rfc793}, are:

\begin{itemize}
    \item \textbf{\LISTEN} represents waiting for a connection request from any remote TCP and port.
    \item \textbf{\SYNSENT} represents waiting for a matching connection request after having sent a connection request.
    \item \textbf{\SYNRECEIVED} represents waiting for a confirming connection request acknowledgement after having both received and sent a connection request.
    \item \textbf{\ESTABLISHED} represents an open connection, data received can be delivered to the user. The normal state for the data transfer phase of the connection.
    \item \textbf{\FINWAITONE{}} represents waiting for a connection termination request from the remote TCP, or an acknowledgement of the connection termination request previously sent.
    \item \textbf{\FINWAITTWO{}} represents waiting for a connection termination request from the remote TCP.
    \item \textbf{\CLOSEWAIT} represents waiting for a connection termination request from the local user.
    \item \textbf{\CLOSING{}} represents waiting for a connection termination request acknowledgement from the remote TCP.
    \item \textbf{\LASTACK{}} represents waiting for an acknowledgement of the connection termination request previously sent to the remote TCP (which includes an acknowledgement of its connection termination request).
    \item \textbf{\TIMEWAIT{}} represents waiting for enough time to pass to be sure the remote TCP received the acknowledgement of its connection termination request.
    \item \textbf{\CLOSED} represents no connection state at all.
\end{itemize}

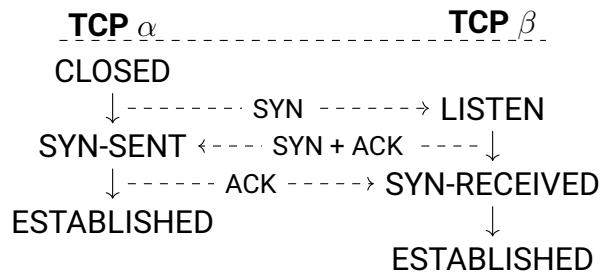
\begin{figure}
    \centering
    \begin{tikzpicture}
        \node (A) at (0, 5) {\bfseries TCP $\alpha$};
        \node (B) at (5, 5) {\bfseries TCP $\beta$};
        \draw[dashed] (-.7,4.8) -- (5.7,4.8);
        \node (TCPA1) at (0, 4.4) {\CLOSED{}};
        \node (TCPB1) at (5, 3.9) {\LISTEN{}};
        \node (TCPA2) at (0, 3.4) {\SYNSENT};
        \node (TCPB2) at (5, 2.9) {\SYNRECEIVED{}};
        \node (TCPA3) at (0, 2.4) {\ESTABLISHED{}};
        \node (TCPB3) at (5, 1.9) {\ESTABLISHED{}};

        \draw[->] (TCPA1) -- (TCPA2);
        \draw[->] (TCPB1) -- (TCPB2);
        \draw[->] (TCPA2) -- (TCPA3);
        \draw[->] (TCPB2) -- (TCPB3);
        \draw[->, dashed] (0.2, 3.9) -- node [midway, fill=white] {\footnotesize SYN} (TCPB1);
        \draw[->, dashed] (4.8, 3.4) -- node [midway, fill=white] {\footnotesize SYN + ACK} (TCPA2);
        \draw[->, dashed] (0.2, 2.9) -- node [midway, fill=white] {\footnotesize ACK} (TCPB2);
    \end{tikzpicture}
    \par
    \caption{Three-way handshake procedure. The solid lines represent the transitions between the states and the dashed lines represent the messages sent.}
    \label{fig:handshake}
\end{figure}

Using the state machine of \Cref{fig:TCPAutomaton}, an example of how a TCP communication can be established between two machines that support the TCP protocol is presented in \Cref{fig:handshake}. When one of the TCP machines, TCP-$\alpha$, is in the \CLOSED{} state  and the other TCP machine, (TCP-$\beta$), is in the state \LISTEN{}, the following steps are needed to establish their TCP connection:

\begin{enumerate}
    \item TCP-$\alpha$ wants to initiate a connection with TCP-$\beta$, thus, it sends a SYN segment to TCP-$\beta$ and moves to the \SYNSENT{} state.

    \item TCP-$\beta$ receives the SYN segment. It has to respond back to TCP-$\alpha$ with a segment containing the SYN and ACK flags, to acknowledge the received segment. Then, it changes its state to \SYNRECEIVED{}.

    \item When TCP-$\alpha$ receives the segment sent by TCP-$\beta$ containing the SYN and ACK flags, it only has to send back an ACK of successfully receiving this segment, and then move to the \ESTABLISHED{} state.

    \item TCP-$\beta$ receives the ACK send by TCP-$\alpha$ and thus also moves to the \ESTABLISHED{} state.
\end{enumerate}

At the end of this procedure, both of the TCP machines are in the \ESTABLISHED{} state and thus they can begin transmitting data. This procedure to open a connection is known as the ``three-way handshake'' in the TCP norm.

\bigskip

For more scenarios on the TCP functionality, such as how two TCP machines can close a connection, the reader can refer to the TCP norm \cite{rfc793}.

\subsubsection{TCP Multitasking}
\label{subsubsec:tcp_mult}

Different tasks can interact and update the socket data structure to handle the various events that can occur within a TCP session. The TCP norm, under Section ``3.9. Event Processing'' page 52, describes a possible implementation of how to handle these events based on three tasks: one for the \emph{user calls}, one for the \emph{arriving segments} and one for the \emph{timers}. This design has also been adopted in \emph{CycloneTCP}. The role of each of the three tasks can be summarized as follow:

\begin{itemize}
\item \textbf{User calls} -- User calls refer to functions, namely OPEN, CLOSE, ABORT, SEND and RECEIVED, that can be called by the user to control the connection, send, or receive data. These functions can trigger transitions between the connection's states since they are intended to control the connection.

\item \textbf{Arriving segments} -- In this task, the received segments are being processed, and the corresponding messages are sent back. Transitions between states can be triggered on the reception of a segment. For example, when a segment containing the SYN flag is received while the socket is in the \LISTEN{} state, a segment with the the flags SYN + ACK will be auto-triggered in response, and the current state will be changed to \SYNRECEIVED{}.

\item \textbf{Timers} -- Timers control the timeouts, for example, the \emph{retransmission} timeout to retransmit a message, or the \emph{time-wait} timeout to close the connection after a specified amount of time elapsed. Thus, corresponding transitions to the timeout events can also be triggered by this task.
\end{itemize}

\begin{figure}
\centering
    \begin{tikzpicture}
        \node [draw, ellipse] (S) at (2, 2) {Socket};
        \node [draw, text width=2.5cm, anchor=north] (T1) at (-2,0) {\textbf{TIMER TASK} \\
                                                Timeout raise};
        \node [draw, text width=2.5cm, anchor=north] (T2) at (2,0) {\textbf{USER TASK} \\
                                                Controlling the connection on the user side};
        \node [draw, text width=2.5cm, anchor=north] (T3) at (6, 0) {\textbf{RECEIVED TASK} \\
                                                Processing received messages};
        \draw[->] (T1) -- (S);
        \draw[->] (T2) -- (S);
        \draw[->] (T3) -- (S);
        \draw[<->, dashed] (T1) -- (T2);
        \draw[<->, dashed] (T2) -- (T3);

        \draw[decorate, decoration={brace, amplitude=10pt}] (7.5, 0) -- (7.5, -2.7)
            node [black,midway,xshift=0.55cm, rotate=-90] {\textit{Tasks}};

        \draw [->] (4.3,2.12) -- (4.9,2.12);
        \node [anchor=west, text width=3.5cm] at (5,2.2) {Socket Operations};
        \draw[<->, dashed] (4.3,1.3) -- (4.9,1.3);
        \node [anchor=west, text width=3.5cm] at (5, 1.3) {Synchronous communications};
    \end{tikzpicture}
    \caption{Concurrency in the TCP protocol, using the 3-tasks based model described in the protocol's specification \cite{rfc793}.}
    \label{fig:TCPConcurrency}
\end{figure}
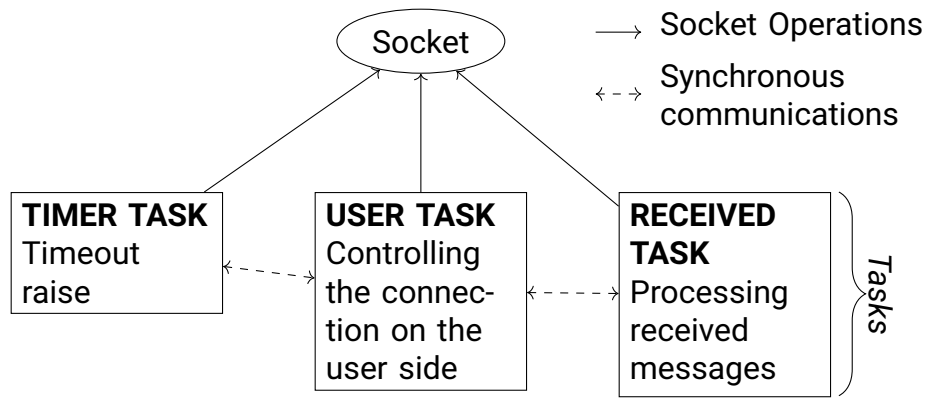

 \Cref{fig:TCPConcurrency} illustrates the interactions between the socket's data structure and the three tasks.  The implementation allows for a single socket to act as a shared resource between the multiple tasks. All tasks can manipulate the socket and change its status concurrently. Only one task can operate on a socket at a given time. The access to the socket's fields is protected by a mutex. Two tasks can communicate synchronously or asynchronously. The synchronous communications are based on the interface provided by the OS, in particular, by the use of events.

\section{The TCP library hardening}

A large part of the TCP behaviour defined by its functional specification in \cite{rfc793} is amenable to automatic proof. For example, formal proof can be used to achieve the:

\begin{itemize}
    \item Verification of the transitions between the different allowed states.
    \item Validation of the integrity of sent messages, \textit{i.e.} we could check that the sent message contains the correct flags and the correct sequence and acknowledgement numbers.
    \item Validation of the integrity of received messages, \textit{i.e.}, we could check that the messages received are correctly processed in regard to the flags they contain.
    \item Functional correctness of the user-task based on the protocols functional specification.
\end{itemize}

The verification of the full functional specification of the \emph{CycloneTCP} library is beyond the scope of this work, as it will require the extensive involvement of the original developers of the library. Instead, our aim is the hardening of the TCP library in the areas that its original authors designated as the most vulnerable or crucial to conform to their functional specifications. Thus, this work mainly focuses on hardening the API of the user task. This mainly falls under two categories: 

\begin{itemize}
\item \textbf{Hardening the user's API} -- One of the most significant problems pointed to by the primary author of the \emph{CycloneTCP} library is the incorrect usage of the library's API. More specifically, the library's users tend to call the API functions in the wrong order, and to forget to check the return code of a function call. Thus, they are subsequently allowing their  TCP implementation to behave outside of the functional specification of the protocol.
\item \textbf{Conformance to the protocol's functional specification} -- The safety of the library largely depends on the assumption that the user's functions are indeed implementing the protocol's specified functionality. Thus, first, we want to verify that the transitions allowed by the current implementation of the user-task related functions respect the state machine of \Cref{fig:TCPAutomaton}, which defines the permitted transitions between the different TCP states. Second, we want to ensure that the user's related functions are always updating a socket's state within the functional specifications of the protocol.
\end{itemize}

So far in the first white paper on the hardening of software libraries with Ada and SPARK, \cite{AdaCore_Hardening_Soft_Libr_EMBENCH}, we demonstrated how the SPARK technology could be used to achieve AoRTE, the Silver level of SPARK assurance. This was done in a bottom-up approach, starting from the lowest level of SPARK assurance and moving to the upper ones, only when the previous level in the hierarchy is completely achieved. As a reminder to the reader, the SPARK level of assurance are:

\begin{enumerate}
\item \textbf{Stone level} -- valid SPARK.
\item \textbf{Bronze level} -- initialization and correct data flow.
\item \textbf{Silver level} -- Absence of Run-Time Errors (AoRTE).
\item \textbf{Gold level} -- proof of key integrity properties.
\item \textbf{Platinum level} -- full functional proof of requirements.
\end{enumerate}

Although a large part of this work was devoted to achieving AoRTE for the code of interest, this process will not be covered in this deliverable. D3.5.7AD offers a wealth of information, examples, and guidelines on how to achieve the Silver level of SPARK assurance, and thus, this will not be repeated in this document. The focus of this deliverable is to step further from proving AoRTE and demonstrate how the SPARK technology can be used to prove key integrity properties and conformance to functional specifications.

The following sections demonstrate the key areas that SPARK hardening was applied and the techniques used to achieve the relevant goals within the two categories of improvements described earlier.

\subsection{Hardening the user's API}

\subsubsection{Enforcing the correct order of API functions}
\label{subsubsec:call_dependencies}

As described earlier in \Cref{subsubsec:tcp_mult}, the user-task implementation offers an API where a user can perform various operations on the network. To achieve this, the API offers several high-level user functions that can alter the state of a socket, which represents the state of the network. The Ada/SPARK functions, translated from their \emph{C} \emph{CycloneTCP} equivalent functions, are located in the files \texttt{socket\_interface.ad(b|s)}. AoRTE was first achieved on these functions, before proceeding with the enforcement of the correct order of calling them.

The TCP protocol implies a specific order such that the user can call the API functions without breaking the functional specification of the protocol. This order also is being conveyed by the protocol's state machine given in \Cref{fig:TCPAutomaton}. Ada's pre- and post-conditions are a powerful tool to express such inter-function dependencies, while SPARK technology can be used to guarantee that the assertions always hold. Thus, post- and pre-conditions were introduced to model a partial order on the calls to the API's functions.

If two functions $f_1$ and $f_2$ are ordered such that $f_1 \preceq f_2$, where $\preceq$ is a relation over the order in which functions have to be called that specifies that $f_1$ must be called before $f_2$, then the post-condition of $f_1$ should be respected by the precondition of $f_2$.

An example of how to enforce such a call-dependency using pre- and post-conditions is given in \Cref{code:FunOrdered} with the procedures \texttt{Socket\_Connect} and \texttt{Socket\_Send}. The procedure \texttt{Socket\_Connect} tries to connect to a distant TCP. If the connection succeeds, the remote IP address of the distant TCP is set in the field \texttt{S\_Remote\_Ip\_Addr} of the \texttt{Sock} structure. When a user calls the procedure \texttt{Socket\_Send}, we want to ensure that the connection has already been established. Thus, the precondition \texttt{Is\_Initialized\_Ip(Sock.S\_Remote\_Ip\_Addr)} of \texttt{Socket\_Send} at line 30 of \Cref{code:FunOrdered}, can only be proved if \texttt{Socket\_Connect} has been called prior to the call of \texttt{Socket\_Send}.

Such proof is enabled by the \emph{Contract\_Cases} aspect at line 8 of the procedure \texttt{Socket\_Connect} in \Cref{code:FunOrdered}. The \emph{Contract\_Cases} aspect will generate a post-condition for the procedure based on the aspect's case that is true upon the entry of the procedure. At each call of the subprogram, one and only one case of the aspect is permitted to evaluate to \emph{true}. The consequence of that case is a contract that is required to be satisfied upon the subprogram's return. Thus, this generated contract acts as a post-condition for the subprogram. In our example, the first case at line 9 checks if the socket is of TCP type, namely \texttt{SOCKET\_TYPE\_STREAM}, at the entry of the procedure. If this is true the consequence of the valid case, namely the \emph{if} statement at line 10--20 will be generated as a post-condition, which is required to be satisfied when the subprogram returns. Thus, this post-condition will provide the automatic provers with the information that the \texttt{S\_Remote\_Ip\_Addr} is correctly set when there is no \texttt{NO\_ERROR}. This information will be used later on to prove the precondition of \texttt{Socket\_Send}, for all calls in the user's code. 

\begin{figure}
    \begin{lstlisting}[language=Ada]
procedure Socket_Connect
    (Sock           : in out Not_Null_Socket;
     Remote_Ip_Addr : in     IpAddr;
     Remote_Port    : in     Port;
     Error          :    out Error_T)
  with
   Pre => Is_Initialized_Ip (Remote_Ip_Addr),
   Contract_Cases => (
     Sock.S_Type = SOCKET_TYPE_STREAM =>
       (if Error = NO_ERROR then
         Sock.S_Type = Sock.S_Type'Old and then
         Sock.S_Protocol = Sock.S_Protocol'Old and then
         Is_Initialized_Ip (Sock.S_localIpAddr) and then
         Sock.S_Local_Port = Sock.S_Local_Port'Old and then
         Sock.S_Remote_Ip_Addr = Remote_Ip_Addr and then
         Sock.S_Remote_Port = Remote_Port and then
         Sock.State = TCP_STATE_ESTABLISHED
       else
         Sock.S_Type = Sock.S_Type'Old and then
         Sock.S_Protocol = Sock.S_Protocol'Old)
     others => True)

procedure Socket_Send
    (Sock    : in out Not_Null_Socket;
     Data    : in     Send_Buffer;
     Written :    out Natural;
     Flags   :        Socket_Flags;
     Error   :    out Error_T)
  with
   Pre  => Is_Initialized_Ip(Sock.S_Remote_Ip_Addr)
 \end{lstlisting}
 \caption{An example of how function calls can be ordered by Pre- and Post-conditions.}
 \label{code:FunOrdered}
\end{figure}

\Cref{tab:call_dependencies} demonstrates all the socket-related user procedures, located in the \texttt{socket\_interface.ad(b|s)} file, that we had to enforce the right call-dependencies using SPARK. The call-dependencies can be determined by the second column which represents the socket's data structure fields that are needed to be set prior to the call of each subroutine, and the third column, which shows which procedures can set these socket's fields. Beyond the dependencies on the socket's fields there are also dependencies on the creation of the socket structure by the \mbox{\emph{Socket\_Open}} procedure. Based on these conditions, the following call-dependencies were enforced using SPARK, similarly to the \Cref{code:FunOrdered} example:

\begin{align*} 
&1.~Socket\textrm{\_}Open \preceq Socket\textrm{\_}Connect\\
&2.~Socket\textrm{\_}Open \preceq Socket\textrm{\_}Send\\
&3.~Socket\textrm{\_}Open \preceq Socket\textrm{\_}Receive\\
&4.~Socket\textrm{\_}Open \preceq Socket\textrm{\_}Shutdown\\
&5.~Socket\textrm{\_}Open \preceq Socket\textrm{\_}Close\\
&6.~Socket\textrm{\_}Connect \preceq Socket\textrm{\_}Send\\
&7.~Socket\textrm{\_}Connect \preceq Socket\textrm{\_}Receive\\
&8.~Socket\textrm{\_}Connect \preceq Socket\textrm{\_}Shutdown
\end{align*}

The transitivity property ensures that when the first dependency is enforced with SPARK then the dependencies 2,3,4 are also enforced by default, due to the dependencies 6,7 and 8.

\begin{table}[]
\centering
\resizebox{\textwidth}{!}{%
\begin{tabular}{|l|l|l|}
\hline
\rowcolor[HTML]{C0C0C0} 
\multicolumn{1}{|c|}{\cellcolor[HTML]{C0C0C0}\textbf{Procedures}} & \multicolumn{1}{c|}{\cellcolor[HTML]{C0C0C0}\textbf{Pre-requirements}} & \multicolumn{1}{c|}{\cellcolor[HTML]{C0C0C0}\textbf{Socket fields set by the procedure or action}} \\ \hline
Socket\_Open & None & Allocation of new socket. No field set. \\ \hline
Socket\_Connect & \begin{tabular}[c]{@{}l@{}}Valid socket.\\ Remote\_Ip\_Addr not set.\end{tabular} & Local\_Ip\_Addr, Local\_Port, Remote\_Ip\_Addr, Remote\_Port. \\ \hline
Socket\_Send &  &  \\ \cline{1-1}
Socket\_Receive &  &  \\ \cline{1-1}
Socket\_Shutdown & \multirow{-3}{*}{\begin{tabular}[c]{@{}l@{}}Valid socket.\\ Remote\_Ip\_Addr set\end{tabular}} & \multirow{-3}{*}{None.} \\ \hline
Socket\_Close & Valid socket. & Free the socket. \\ \hline
\end{tabular}%
}
\caption{The socket-related user procedures, located in the \texttt{socket\_interface.ad(b|s)} file, that their call-dependencies must be met. The call-dependencies can be determined by the second column which represents the socket's data-structure fields that are needed to be set prior to the call of each subroutine, and the third column, which shows which procedures can set these socket's fields. Beyond the dependencies on the socket's fields there are also dependencies on the creation of the socket structure by the \emph{Socket\_Open} procedure.}
\label{tab:call_dependencies}
\end{table}

\subsubsection{Checking the correctness of return codes}

An observation made by the development team of the \emph{CycloneTCP} library is that their customers often neglect to check the return code of the socket-related user functions. Thus, they do not deal with the possibility of function-calls failing to achieve their goal and returning an error code. If the return code is not checked, no assumption can be made on the validity of the execution that follows a function call which has the potential of producing an error.

Post-conditions can be used to enforce checks on the return code of a procedure. Such checks will be required after the call of this procedure by GNATprove to enable proving of the program. To understand the mechanisms of this approach, one must understand GNATprove's goal. GNATprove aims to compute all possible values of variables to guarantee the AoRTE. Thus, the user has to either provide extra information to GNATprove or write the code in such a way that eliminates any ambiguity about a variable's state. GNATprove will generate a warning if it can not deduct the possible states of a variable at any place in the code. In \Cref{subsubsec:call_dependencies}, we explained how the \emph{Contract\_Cases} aspect used in \Cref{code:FunOrdered} at line 8 will generate a post-condition for the \texttt{Socket\_Connect} procedure. This generated post-condition will convey the following information after each call of the \texttt{Socket\_Connect}:

\begin{enumerate}
    \item if the \texttt{ERROR} is equal to \texttt{NO\_ERROR} the TCP session will be in the \texttt{TCP\_STATE\_ESTABLISHED} state, and all the relevant socket's data structure fields will be set, including \texttt{S\_Remote\_Ip\_Addr}, otherwise
    \item the field \texttt{S\_Remote\_Ip\_Addr} can have any value.
\end{enumerate}

Thus, if the user calls the two procedures of \Cref{code:FunOrdered} in the following manner:

\begin{lstlisting}[language=Ada]
Socket_Connect (Sock, Remote_Ip_Addr, Port, Error);
Socket_Send (Sock, Data, Written, Flags, Error);
\end{lstlisting}

GNATprove will warn that:

\begin{verbatim}
medium: precondition might fail.
\end{verbatim}

This is because, depending on the \texttt{ERROR} value, the introduced post-condition will provide the information to GNATprove that \texttt{Sock.S\_Remote\_Ip\_Addr} is either set or not. Thus, when a call to the \texttt{Socket\_Send} procedure follows, its pre-condition, given at line 30 of \Cref{code:FunOrdered}, will fail because of the ambiguity on the value of \texttt{Sock.S\_Remote\_Ip\_Addr}. The user is now enforced to check the return code of the \texttt{Socket\_Connect} after the function returns to resolve the ambiguity on the TCP's session state: 

\begin{lstlisting}[language=Ada]
Socket_Connect (Sock, Remote_Ip_Addr, Port, Error);
if Error /= NO_ERROR then
   -- TCP session state -> no connection established
   Socket_Close(Sock);
else
   -- TCP session state -> connection established
   Socket_Send (Sock, Data, Written, Flags, Error);
   -- continue processing
end if;
\end{lstlisting}

Now the prover knows that \texttt{Socket\_Send} will only be called when \texttt{Socket\_Connect} has returned without an error code, and thus, the \texttt{Socket\_Send}'s pre-condition can be proved. In the case of \texttt{Socket\_Connect} returning an error code, we decided to close the socket and bring the TCP session to the \texttt{CLOSED} state. Users can handle this case more precisely depending on the exact error-code returned, for example, \texttt{Error = ERROR\_PORT\_UNREACHABLE}, and provide their implementation of how to proceed in each error case; always respecting the TCP specification.

\subsection{Conformance to the protocol's functional specification}
\label{subsec:functionalSpecs}

The original \emph{CycloneTCP} library provides no guarantees that the implementation respects the TCP functional specification. Thus, there is no assurance that the implementation will only allow valid transitions between the different states that a TCP session can exhibit. We aim to use the SPARK technology to verify that the implementation respects the TCP automaton, given in \Cref{fig:TCPAutomaton}. Although, this work is focused on hardening the API of the user task, mainly the high-level user functions located in the file \texttt{tcp\_interface.ad(b|s)}, to verify the state transitions the rest of the TCP protocol has to be taken into account. The reason for this is two-fold. First, although user-functions can trigger state transitions, the actual transitions are done by library functions that do not belong to the user-task set of functions. Second, other parts of the library can trigger state transitions during and between the user-task function calls, which can affect the intermediate and final states that a user related function can exhibit. 

The TCP user API related functions were fully translated to SPARK. SPARK bindings were also provided to the majority of the rest of the library's C functions to allow their invocation from the SPARK code. Furthermore, SPARK helper functions were introduced to assist with the state-transition validation. \Cref{tab:SPARK_code} gives a list of all the files we needed to introduce or alter, to complete this work. The table also provides a short description of their purpose. 

\begin{table}[]
\centering
\resizebox{\textwidth}{!}{%
\begin{tabular}{|l|l|l|}
\hline
\rowcolor[HTML]{C0C0C0} 
\multicolumn{1}{|c|}{\cellcolor[HTML]{C0C0C0}\textbf{File}} & \multicolumn{1}{c|}{\cellcolor[HTML]{C0C0C0}\textbf{Description}} & \multicolumn{1}{c|}{\cellcolor[HTML]{C0C0C0}\textbf{Translation or binding}} \\ \hline
ada\_main.adb/s & Customer SPARK code that tests the socket API. & SPARK code. \\ \hline
socket\_types.ads & Types and structure of a socket. & SPARK code \\ \hline
socket\_interface.adb/s & Socket API. & SPARK code. \\ \hline
socket\_helper.ads & Helper function for proofs. & Helper functions for proofs. \\ \hline
tcp\_type.ads & Types used for TCP. & SPARK code. \\ \hline
tcp\_interface.adb/s & TCP user functions. & SPARK code. \\ \hline
tcp\_misc\_binding.adb/s & Helper functions for TCP. & SPARK code / binding to C code. \\ \hline
tcp\_fsm\_binding.ads & TCP finite state machine. Functions to process incoming segments. & Binding to C code. \\ \hline
tcp\_timer\_interface.ads & Simulate a timer tick. & SPARK code. Helper functions for proofs. \\ \hline
udp\_binding.adb/s & UDP functions. & Binding to C code. \\ \hline
net\_mem\_interface.adb/s & Memory management. & Binding to C code. \\ \hline
ip\_binding.adb/s & Underlaying IP layer functions. & Binding to C code. \\ \hline
\end{tabular}%
}
\caption{}
\label{tab:SPARK_code}
\end{table}

\subsubsection{Technical challenges encountered and overview of solutions}

The validation of functional specifications with SPARK can be a quite challenging task depending on the complexity, the size, and the implementation of the project. In the case of the \emph{CycloneTCP} TCP implementation, the following two issues required innovative solutions:

\subsubsubsection{Modelling concurrent behaviour with SPARK}

The concurrent implementation of the TCP protocol allows for multiple interactions between the tasks, synchronous or asynchronous, and thus, numerous possible state changes at any given moment. This makes the functional-specification verification significantly challenging, as SPARK does not have a native mode to deal precisely with interactions related to concurrency. To address this, we introduced SPARK functions with contracts that model how the different possible interactions can modify the state of a socket. The solution will be elaborated in \Cref{subsubsec:dealing_concurrency}. 

\subsubsubsection{Verifying non-user functions}
\phantomsection
\label{verifying_non-user_funcs}

The TCP/IP specification document provides all the information needed for the allowed transitions between the different TCP possible states. For any transitions that are directly triggered through the call of the user-tasks functions, SPARK contracts, that represent transitions allowed by the functional specification, are embedded within the SPARK translated code. Then \emph{GNATprove} can be used to prove that the code will always respect those transitions. The SPARK contracts were placed into the \texttt{Tcp\_Change\_State} procedure, found in the SPARK translated file \texttt{tcp\_misc\_binding.ads}. The \texttt{Tcp\_Change\_State} is a helper function that is called every time a state transition needs to be made, to update the socket's state. An incorrect transition allowed by the code will be detected by the prover.

The above approach can not be used in the case of transitions that are not related to the user-task. This is because the code of these functions is not translated to SPARK, but rather only SPARK bindings to the C code are introduced. Thus the prover can not be used to evaluate the validity of the state transitions allowed by the C code.

The first step of verifying the non-user related state transitions is to identify their source. As shown in \Cref{fig:TCPAutomaton}, the majority of these state transitions are triggered through segment transmissions. The \texttt{tcp\_fsm.c} file includes all the functions that are responsible for processing received segments. The main function of this file \texttt{tcpProcessSegment} looks for the socket corresponding to the received segment, and then according to the current TCP state of this socket, it calls one of the \texttt{tcpState<StateName>} functions to process the segment. This is shown the the following code taken from the \texttt{tcpProcessSegment} function:

\begin{lstlisting}[language=C]
//Check current state
switch(socket->state)
{
//Process CLOSED state
case TCP_STATE_CLOSED:
  //This is the default state that each connection starts in before
  //the process of establishing it begins
  tcpStateClosed(interface, pseudoHeader, segment, length);
  break;
//Process LISTEN state
case TCP_STATE_LISTEN:
  //A device (normally a server) is waiting to receive a synchronize
  //(SYN) message from a client. It has not yet sent its own SYN message
  tcpStateListen(socket, interface, pseudoHeader, segment, length);
  break;
//Process SYN_SENT state
case TCP_STATE_SYN_SENT:
  //The device (normally a client) has sent a synchronize (SYN) message
  //and is waiting for a matching SYN from the other device (usually
  //a server)
  tcpStateSynSent(socket, segment, length);
  break;
//Process SYN_RECEIVED state
case TCP_STATE_SYN_RECEIVED:
  //The device has both received a SYN from its partner and sent its own
  //SYN. It is now waiting for an ACK to its SYN to finish connection
  //setup
  tcpStateSynReceived(socket, segment, buffer, offset, length);
  break;
//Process ESTABLISHED state
case TCP_STATE_ESTABLISHED:
  //Data can be exchanged freely once both devices in the connection
  //enter this state. This will continue until the connection is closed
  tcpStateEstablished(socket, segment, buffer, offset, length);
  break;
//Process CLOSE_WAIT state
case TCP_STATE_CLOSE_WAIT:
  //The device has received a close request (FIN) from the other device.
  //It must now wait for the application to acknowledge this request and
  //generate a matching request
  tcpStateCloseWait(socket, segment, length);
  break;
//Process LAST_ACK state
case TCP_STATE_LAST_ACK:
  //A device that has already received a close request and acknowledged
  //it, has sent its own FIN and is waiting for an ACK to this request
  tcpStateLastAck(socket, segment, length);
  break;
//Process FIN_WAIT_1 state
case TCP_STATE_FIN_WAIT_1:
  //A device in this state is waiting for an ACK for a FIN it has sent,
  //or is waiting for a connection termination request from the
  //other device
  tcpStateFinWait1(socket, segment, buffer, offset, length);
  break;
//Process FIN_WAIT_2 state
case TCP_STATE_FIN_WAIT_2:
  //A device in this state has received an ACK for its request to
  //terminate the connection and is now waiting for a matching FIN
  //from the other device
  tcpStateFinWait2(socket, segment, buffer, offset, length);
  break;
//Process CLOSING state
case TCP_STATE_CLOSING:
  //The device has received a FIN from the other device and sent an ACK
  //for it, but not yet received an ACK for its own FIN message
  tcpStateClosing(socket, segment, length);
  break;
//Process TIME_WAIT state
case TCP_STATE_TIME_WAIT:
  //The device has now received a FIN from the other device and
  //acknowledged it, and sent its own FIN and received an ACK for
  //it. We are done, except for waiting to ensure the ACK is
  //received and prevent potential overlap with new connections
  tcpStateTimeWait(socket, segment, length);
  break;
//Invalid state...
default:
  //Back to the CLOSED state
  tcpChangeState(socket, TCP_STATE_CLOSED);
  //Silently discard incoming packet
  break;
}
\end{lstlisting}

 The family of \texttt{tcpState<StateName>} functions, also located in the \texttt{tcp\_fsm.c} file, check the information contained in the segment and can perform a change of state in the socket structure depending on the segment's flags. Any of these functions that have to perform a state transition has a call to the \texttt{Tcp\_Change\_State} procedure which will perform the transition. 

Trying to extract all the possible state transitions allowed by the \emph{CyloneTCP} TCP implementation by manual code inspection is an expensive and error-prone process. One way to approach this is to locate all the calls to the \texttt{Tcp\_Change\_State} procedure and try to understand the state transitions performed at each one of these cases. Instead, we decided to deploy automatic techniques to extract the possible state transitions triggered by non-user-task related functions. This is where SPARK technology excels. Even though the C part of the code can't be proved, it allows for a hybrid-verification approach, where other traditional testing approaches can be used to test that the behaviour of the code matches the functional specifications. SPARK procedures that model the state transitions of the C code can then be enhanced with contracts that reflect this verified behaviour. These new contracts allow GNATprove to verify any parts of the SPARK code that have dependencies on those C bindings. This allows for the hardening of software libraries that are not completely translated to SPARK. The hybrid approach used for this work is based on symbolic execution and is being explained in \Cref{subsubsect:KLEE}.

\subsubsection{Dealing with concurrency}
\label{subsubsec:dealing_concurrency}

The TCP concurrent implementation of the \emph{CycloneTCP} library is based on the \emph{mutex} synchronization mechanism. The socket of a TCP connection is protected by a \emph{mutex}, namely the \texttt{netMutex}, and only one function at any given time can lock this mutex and perform changes to the socket's state. Interactions must be considered at two locations: between the function calls, such as the \texttt{Open} or \texttt{Close} user functions, or during a function call, when the program waits for an event, such as \texttt{rcv} (see \Cref{fig:TCPAutomaton}). In both cases, the mutex on the socket is being released. Sections \ref{section:asynchronous} and \ref{section:synchronous} deal with these two cases, respectively.

\subsubsubsection{Concurrency: asynchronous changes of state}
\label{section:asynchronous}

Between the user-task related function calls, the mutex that protects the socket structure is released, and segments can be received. The reception of a segment can lead to socket state changes, which represents TCP state transitions, and are considered asynchronous to the user-task related functions. Between two function calls, an infinite number of segments can be potentially received. To model any possible state-changes upon the reception of a segment in SPARK, the function \lstinline[language=Ada]{Tcp_Process_One_Segment} is introduced. The iteration over the \lstinline[language=Ada]{Tcp_Process_One_Segment}, when multiple segments are received, must also be considered to compute the resulting state after the iteration completes.

Let $\rightarrow$ be the function modelling the transitions of the TCP automaton, restricted to the possible transitions that can be performed by the reception of a message and its automatic response mechanism. Thus in our case, $\rightarrow$ represents the \text{\lstinline[language=Ada]{Tcp_Process_One_Segment}} function. We also need to consider the reflexive transitive closure \cite{wiki:transitive_closure} $\rightarrow^*$ of $\rightarrow$, with

 \[\rightarrow^* = \bigcup_{n\in\mathbb{N}} \rightarrow^n\]

\noindent
where $\bigcup$ represents the final result over calling the \text{\lstinline[language=Ada]{Tcp_Process_One_Segment}} repeatedly. By examining the TCP automaton in \Cref{fig:TCPAutomaton}, the maximum number of automatic state transitions triggered by received messages is three. This represents the longest path between transitions where its edges are correlated to a \emph{rcv} action; namely the maximum path is between the \SYNSENT{} and \CLOSEWAIT{}, going through the \SYNRECEIVED{} and \ESTABLISHED{} states without any user-related action. Thus, since we only consider the transitive closure in terms of states, and not in the number of continuously received segments, we can significantly reduce the number of iterations of $\rightarrow$ to compute $\rightarrow^*$:

\[\rightarrow^* = \bigcup_{n=1}^3 \rightarrow^n\]

The algorithm given in \ref{algo:tcpProcessSegment} is sufficient to compute the transitive closure of the function \lstinline[language=Ada]{Tcp_Process_One_Segment} by taking advantage of the fact that SPARK can unroll small loops to enable the proving. %The contract of \lstinline[language=Ada]{Tcp_Process_Segment} is manually written and proved with SPARK.

\begin{algorithm}[t]
    \SetKwFunction{ProcessSegm}{\lstinline[language=Ada]{Tcp\_Process\_Segment}}
    \SetKwProg{Fn}{function}{}{end}
    \Fn{\ProcessSegm{Socket}}{
        $S_{last} := \text{\textit{Socket}}$\;
        $S := S_{last}$\;
        \For{$i=1$ \KwTo $3$}{
            $S_{last} :=$ \lstinline[language=Ada]{Tcp_Process_One_Segment}($S_{last}$) \;
            $S := S \cup S_{last}$\;
        }
        \KwRet{$S$}\;
    }
    \caption{Algorithm to compute the reflexive and transitive closure of \lstinline[language=Ada]{Tcp_Process_One_Segment}}
    \label{algo:tcpProcessSegment}
\end{algorithm}

A call to the \lstinline[language=Ada]{Tcp_Process_Segment} function is added everywhere there is a call to \lstinline{Os_Acquire_Mutex (Net_Mutex)} to consider all the possible states that the TCP connection can be in at the point of calling a user function. The choice of the \lstinline[language=Ada]{Tcp_Process_Segment} function-calls placement is based on the fact that the \lstinline{Os_Acquire_Mutex (Net_Mutex)} is called at the beginning of the user functions, and after its call, the mutex is held by the user function. Thus, no other thread will be able to interfere and change the TCP state after the user function has acquired the mutex.

\subsubsubsection{Concurrency: synchronous exchange}
\label{section:synchronous}

The second case in which the resulting state of a user function can be affected by the rest of the TCP tasks is when the user's functions release the mutex to allow for other events to take place. In this case, the C function \lstinline[language=C]{Tcp_Wait_For_Events} is called from the user function to check if the event requested is completed. For example, the \texttt{Tcp\_Connect} user function, located in the \texttt{tcp\_interface.adb}, calls the \lstinline[language=C]{Tcp_Wait_For_Events} to wait for the completion of the socket connection event, namely the \texttt{SOCKET\_EVENT\_CONNECTED}. The call to the \lstinline[language=C]{Tcp_Wait_For_Events} releases the mutex, and when the \emph{arriving} task receives the appropriate segments, in this case, a segment with the \texttt{SYN} flag, the connection will be established. This will move the socket to the \ESTABLISHED{} state. In the meantime, the \texttt{Tcp\_Update\_Events} function is monitoring for any state changes and when it notices that the connection is established it will trigger the \texttt{SOCKET\_EVENT\_CONNECTED} event, using the OS event mechanism. This will allow the \texttt{Tcp\_Connect} user function to resume execution.

Essentially, the function \lstinline[language=C]{Tcp_Wait_For_Events}, located in the \texttt{tcp\_misc.c} file releases the mutex to allow for required events to happen. The function \lstinline[language=C]{Tcp_Update_Events}, located in the same file, detects that the event expected outcome is completed by monitoring the changes to the TCP states. Then it updates the specified event to be true in the socket structure, and it raises the desired event to allow the resuming of the user function that requested the event in the first place. Thus, the \lstinline[language=C]{Tcp_Update_Events} function is called when a segment is being received, and a state change took place. This makes the \lstinline[language=C]{Tcp_Update_Events} function a perfect candidate to introduce contracts in SPARK and model the possible states after the completion of each event.

Algorithm \ref{algo:waitForEvents}, which reuses \lstinline[language=Ada]{Tcp_Process_One_Segment}, computes the set of possible final states after the completion of an event. This is implemented by the \lstinline[language=Ada]{Tcp_Wait_For_Events_Proof} function, located in the \texttt{tcp\_misc\_binding.adb} file. This function is dedicated to proving the conformance to the functional specifications of the protocol in regards to the possible states that the TCP session can exhibit after each event completion.

\begin{algorithm}[t]
    \SetKwFunction{ProcessSegm}{\lstinline[language=Ada]{Tcp\_Wait\_For\_Events_Proof}}
    \SetKwProg{Fn}{function}{}{end}
    \Fn{\ProcessSegm{Socket, Event, Event\_Mask}}{
        $S_{last} := \text{\textit{Socket}}$\;
        $S := S_{last}$\;
        $E :=$ \lstinline[language=Ada]{Tcp_Update_Events}($S_{last}$)\;
        \If{$(E\ \&\ \text{Event\_Mask}) \neq 0$}{
            \KwRet{$S$}\;
        }
        \For{$i=1$ \KwTo $3$}{
            $S_{last} :=$ \lstinline[language=Ada]{Tcp_Process_One_Segment}($S_{last}$) \;
            $S := S \cup S_{last}$\;
            $E :=$ \lstinline[language=Ada]{Tcp_Update_Events}($S_{last}$)\;
            \If{$(E\ \&\ \text{Event\_Mask}) \neq 0$}{
                \KwRet{$S$}\;
            }
        }
        \KwRet{$\emptyset$}\;
    }
    \caption{Function to compute the possible state after the completion of a particular event that is requested by a user-task related function.}
    \label{algo:waitForEvents}
\end{algorithm}

We can compute precisely the states reached for each expected event thanks to the fact that SPARK unrolls loops.

\subsubsection{Enhancing the library's security with symbolic execution and SPARK}
\label{subsubsect:KLEE}

Although SPARK does not have a native-mechanism to deal with concurrency, in  \Cref{subsubsec:dealing_concurrency} we demonstrated how such a mechanism could be improvised to allow the modelling of concurrency and all the possible interactions that can affect the state of a TCP connection using SPARK. More specifically, the three introduced functions, \text{\lstinline[language=Ada]{Tcp_Process_One_Segment}}, \lstinline[language=Ada]{Tcp_} \lstinline[language=Ada]{Process_Segment}, and \lstinline[language=Ada]{Tcp_Wait_For_Events_Proof}, are able to model the state changes that happen in functions that are not related to the user-task. Thus, these functions can be used to prove that the allowed state transitions of the \emph{CycloneTCP} implementation conform to the functional specifications of the TCP protocol. The next step to enable this is to add appropriate contracts to the introduced functions. These contracts will reflect only the state transitions documented by the protocol’s specification.

In the case of the user-task related functions, introducing and proving such contracts is straightforward as the whole code was translated to SPARK. In the case of non-user-related tasks, any introduced contracts that are needed to prove the validity of state transitions can not be proved by SPARK since the code is only written in C. To overcome this issue, a hybrid approach can be deployed, where symbolic execution is used to test the conformance of the transitions allowed by the implementation using assertions and symbolic execution. The assertions introduced in the C code for testing are based on the protocol's functional specification in \cite{rfc793}. Then, when assurances are gained through symbolic execution that the C code is conforming to the protocols functional specifications, the tested assertions are translated to SPARK contracts and added to the introduced functions that model the state transitions, namely, \text{\lstinline[language=Ada]{Tcp_Process_One_Segment}}, \lstinline[language=Ada]{Tcp_Process_Segment}, and \lstinline[language=Ada]{Tcp_Wait_For_Events_Proof}. The new assertions are then used to enable the proving of the user API code.

\subsubsubsection{Symbolic execution brief introduction}

Symbolic execution is a technique introduced to overcome the shortcomings of traditional testing approaches. In most realistic applications, the input space is too large to be exhaustively tested by any traditional testing method. Let's consider the case of testing a program using randomly generated inputs to identify if the program's implementation is violating any of the functional specifications and leading to security violations. Since random testing usually covers only a fraction of the possible input space, it is highly possible for the testing to miss important input cases that reveal such security vulnerabilities. Symbolic execution provides an alternative solution to this problem. Rather than using concrete inputs to test a program, it represents the program's inputs abstractly as symbols \cite{pasareanu2020symbolic}. Then it progresses the execution symbolically constructing: a) expressions based on the symbols and taking into account the expressions and variables found on the symbolically executed path, and b) constraints in terms of those symbols and by accounting for each branch statement's possible outcomes. Solving these expressions and constraints using constraint solvers allows for the construction of test cases that could cause the violation of the property under investigation (if such cases exist).

Although exhaustive symbolic execution can be considered sound and complete (it prevents false negatives and false positives \cite{symbolic_exec_survey:2018}), and thus, all the possible unsafe inputs can be captured, in reality, the technique suffers from scalability issues related mainly to state space explosion. Therefore, such an approach can not, in practice, replace the use of formal verification. Nevertheless, in this work, we trade soundness for performance to gain further software assurances. Although this hybrid-approach can not be considered complete because the C code has not been tested exhaustively with symbolic execution, it can still significantly contribute to the hardening of libraries that are not fully converted to SPARK. The approach provided confidence that state transitions, triggered by non-user-task related functions but capable of affecting the user task's state, are valid.

For this work the \emph{KLEE}\footnote{\url{https://klee.github.io}} symbolic execution engine was used. To better understand the work done with \emph{KLEE}, an example of its usage within the scope of this work is given in the following section.

\subsubsubsection{Example of gaining assurance on the conformance to the protocol's functional specification using \emph{KLEE}}

As explained in Section 4.2.1.2, the function \texttt{tcpProcessSegment}, (located in the file \texttt{tcp\_fsm.c}) is responsible for processing the incoming segments depending on the state of the socket that received the segment. As shown in the code given in Section 4.2.1.2, for each of the possible cases of states, one of the \texttt{tcpState<StateName>} functions will be invoked from \texttt{tcpProcessSegment} to handle the segment. A call to one of the functions can cause a state of change, which will be reflected in the socket structure. Thus the \emph{KLEE} symbolic execution can be utilized to test if each one of these functions respects the states transitions specified by the protocol's functional specification.

Let's take for example the \texttt{tcpProcessCloseWait}, function from the \texttt{tcpState} \texttt{<StateName>} family of functions. This function is called from the \texttt{tcpProcessSegment} function when a segment is received and the socket is in the state \texttt{CLOSEWAIT}. The TCP protocol's functional specification describes the expected behaviour as follows:

\begin{itemize}
\item If the segment contains the \texttt{RST} bit (Reset flag) then the TCP connection state becomes \texttt{CLOSED}.
\item No other transition can be done, since the only transition to \texttt{LASTACK} requires an action by the user.
\end{itemize}

Thus, we want to prove that the C code implementation respects these rules extracted from the TCP specifications. To achieve this, symbolic execution is used to verify that all the possible paths when executing the \texttt{tcpProcessCloseWait} function will result in a state that respects the above rules. This process requires four steps:

\begin{enumerate}
\item Using the \emph{KLEE} symbolic execution framework, incoming segments are represented symbolically. This representation allow us to account for all possible incoming segments.
\item The state of the socket is changed to \texttt{CLOSEWAIT} to force the execution to go through the \texttt{tcpProcessCloseWait} function.
\item The function \texttt{tcpProcessCloseWait} is executed symbolically with KLEE.
\item After the call to the \texttt{tcpProcessCloseWait} function, a \texttt{klee\_assert} is introduced that checks if the code leads to a valid state identified in the protocol's specification; in this case the two specifications given above for the possible outcome of calling the \texttt{tcpProcessCloseWait} function. The following C code demonstrates the implementation of this four steps:
\end{enumerate}

\begin{lstlisting}[language=C]
int main() {
    // Initialisation
    socketInit();
    Socket* socket;
    TcpHeader *segment;
    size_t length;
    uint32_t ackNum, seqNum;
    uint8_t flags;
    uint16_t checksum;
    struct socketModel* sModel;
    segment = malloc(sizeof(TcpHeader));
    // creation of a TCP socket
    socket = socketOpen(SOCKET_TYPE_STREAM, SOCKET_IP_PROTO_TCP);

    // Step 1 : Representing the incoming segment symbolically
    klee_make_symbolic(&ackNum, sizeof(ackNum), "ackNum");
    klee_make_symbolic(&seqNum, sizeof(seqNum), "seqNum");
    klee_make_symbolic(&flags, sizeof(flags), "flags");
    klee_assume(flags >= 0 && flags <= 31);
    klee_make_symbolic(&checksum, sizeof(checksum), "checksum");
    segment->srcPort = 80;
    segment->destPort = socket->localPort;
    segment->seqNum = seqNum;
    segment->ackNum = ackNum;
    segment->reserved1 = 0;
    segment->dataOffset = 6;
    segment->flags = flags;
    segment->reserved2 = 0;
    segment->window = 26883;
    segment->checksum = checksum;
    segment->urgentPointer = 0;
    klee_make_symbolic(&length, sizeof(length), "length");

    // Step 2 : Change of the socket state.
    tcpChangeState(socket, TCP_STATE_CLOSE_WAIT);
    klee_assert(socket->state == TCP_STATE_CLOSE_WAIT);
    // We make a hard copy of the struct to check if the fields
    // representing the Model of the socket have changed after
    // the call to tcpStateCloseWait - a change would represent
    // a potential bug in this case.
    sModel = toSockModel(socket);

    // Step 3 : Execution of the function
    tcpStateCloseWait(socket, segment, length);
    }
    // Step 4 : Verification of the assertion that represents the allowed
    // by the protocol's specification states to be reached, after the
    // execution of the function in question; in this case the
    // tcpStateCloseWait function
    klee_assert(equalSocketModel(socket, sModel) &&
               (socket->state == TCP_STATE_CLOSE_WAIT ||
               (socket->state == TCP_STATE_CLOSED &&
                socket->resetFlag)));
\end{lstlisting}

Using the above code, \emph{KLEE} runs exhaustive symbolic execution on the \lstinline[language=Ada]{tcpStateCloseWait} function without raising an error on the introduced assertion. This proves that the function respects the TCP functional specification. Thus, the assertion used in step four of the above example can be translated into a SPARK post-condition for the \lstinline[language=Ada]{Tcp_Process_One_Segment} function. As explained earlier in \Cref{subsubsec:dealing_concurrency}, this function is essential for the modelling and the verification of concurrency and is used to prove that the concurrent interactions of the user-task with the rest of the tasks lead to valid TCP states; states that respect the protocol's functional specification. The code's security relies on the confidence we have on the \lstinline[language=Ada]{Tcp_Process_} \lstinline[language=Ada]{One_Segment} function. Thus, the symbolic execution approach followed for the \lstinline[language=Ada]{tcpStateCloseWait} example is used for all the functions of the same family, namely functions with the format \lstinline[language=Ada]{tcpState<StateName>}. In each case, if the protocol's specification is respected, then there is no error raised by the symbolic execution and the relevant assertion is also introduced as a post-condition to \lstinline[language=Ada]{Tcp_Process_One_Segment}. In the case that the symbolic execution raises an error, the code does not respect the protocol's specification, and this is considered a bug to be fixed. An example, of such detected bug, is given in the following section.

\subsection{Bug found and resolved}

Thanks to the SPARK work done for proving the conformance of the implementation to the TCP protocol's functional specifications, a bug was found in the original C implementation. In general, when a direct translation from C to SPARK source code is applied, there is always the risk that the SPARK code will inherit any diversions from the functional specifications that the C code already has. Any such diversions from the functional specification can be captured by expressing the program's functional specifications with SPARK contracts and by trying to prove them.

In our case, the initial translated SPARK user-task functions were also based on the C code, and thus, their functional correctness could not be taken for granted. Therefore, we introduced the correct contracts that represent the TCP's functional specifications on the appropriate procedures, such as \texttt{Tcp\_Change\_State} (see \Cref{subsec:functionalSpecs}), and we used the SPARK technology to prove them. As explained in \Cref{subsec:functionalSpecs}, such SPARK contracts represent all the valid transitions between TCP states described by the TCP automaton in \Cref{fig:TCPAutomaton}. Thus any attempt to prove these contracts while the source code does not conform to them will result in the SPARK provers raising a warning. Such a warning means that the TCP protocol's current implementation has a deviation from the protocol's functional specification, and thus, a bug exists in the implementation.

Indeed, the contract placed on the \lstinline[language=Ada]{Tcp_Change_State} procedure to represent the permitted by the protocol's functional specifications state transitions was able to capture a bug of an unauthorized state transition. The bug was found in the \texttt{Tcp\_shutdown} procedure of the \texttt{Tcp\_interface.adb} file. The part of the \texttt{Tcp\_shutdown} procedure's code that contained the bug is listed below:

\begin{lstlisting}[language=Ada, basicstyle=\ttfamily\small, numbers=left, numberstyle=\tiny, escapechar=\%]
case Sock.State is
  when TCP_STATE_SYN_RECEIVED
     | TCP_STATE_ESTABLISHED =>
  -- Flush the send buffer
  Tcp_Send (Sock, Buf, Ignore_Written,
            SOCKET_FLAG_NO_DELAY, Error);
  if Error /= NO_ERROR then
     return;
  end if;

  -- Make sure all the data has been sent out
  Tcp_Wait_For_Events % \label{bugProg:tcpWaitForEvents} %
     (Sock       => Sock,
      Event_Mask => SOCKET_EVENT_TX_DONE,
      Timeout    => Sock.S_Timeout,
      Event      => Event);

  -- Timeout error?
  if Event /= SOCKET_EVENT_TX_DONE then
     Error := ERROR_TIMEOUT;
     return;
  end if;

  -- Send a FIN segment
  Tcp_Send_Segment   % \label{bugProg:tcpWaitForEventsPost} %
     (Sock         => Sock,
      Flags        => TCP_FLAG_FIN or TCP_FLAG_ACK,
      Seq_Num      => Sock.sndNxt,
      Ack_Num      => Sock.rcvNxt,
      Length       => 0,
      Add_To_Queue => True,
      Error        => Error);
  -- Failed to send FIN segment?
  if Error /= NO_ERROR then
     return;
  end if;
  -- Switch to the FIN-WAIT-1 state
  Tcp_Change_State (Sock, TCP_STATE_FIN_WAIT_1); % \label{bugProg:tcpChangeState} %
\end{lstlisting}

The function \lstinline[language=Ada]{Tcp_Send} called at line 5 in the above code snippet changes the state of the socket to either
the \ESTABLISHED{} or \CLOSEWAIT{} if no error exists. The procedure \texttt{Tcp\_Wait\_For\_Events} is then called at line \ref{bugProg:tcpWaitForEvents}. During this call, the mutex that guards the socket structure can be released, and at this point, there is a possibility that the remote can reset the connection. In this scenario, the socket's state is changed to either the \ESTABLISHED{}, \CLOSEWAIT{}, or \CLOSED{} state. This was the C code's original behavior, which was also translated to SPARK and encoded to the post-condition contract that was originally introduced to the \texttt{Tcp\_Wait\_For\_Events} procedure. The relevant part of this contract is given below:

\begin{lstlisting}[language=Ada, basicstyle=\ttfamily\small]
(if (Event_Mask and SOCKET_EVENT_TX_DONE) /= 0 then
(if Event = SOCKET_EVENT_TX_DONE then
  -- RST segment received
  Model(Sock) = (Model(Sock)'Old with delta
     S_State => TCP_STATE_CLOSED,
     S_Reset_Flag => True) or else
  (if Sock.S_State'Old = TCP_STATE_CLOSE_WAIT then
     Model(Sock) = Model(Sock)'Old or else
  elsif Sock.S_State'Old = TCP_STATE_ESTABLISHED then
     Model(Sock) = Model(Sock)'Old or else
     Model(Sock) = (Model(Sock)'Old with delta
        S_State => TCP_STATE_CLOSE_WAIT))))
\end{lstlisting}

Thus, in case the above scenario is executed after the call of the \texttt{Tcp\_Wait\_For\_Events} procedure at line \ref{bugProg:tcpWaitForEvents}, the execution continues to line \ref{bugProg:tcpWaitForEventsPost} where the \lstinline[language=Ada]{ Tcp_Send_Segment} is called to send a \texttt{FIN} segment, and thus, the state transitions \CLOSEWAIT{} $\rightarrow$ \FINWAITONE{}
 and \CLOSED{} $\rightarrow$ \FINWAITONE{} are possible. Then at line \ref{bugProg:tcpChangeState} the procedure \lstinline[language=Ada]{Tcp_Change_State} is called to perform the possible transition of states. As explained in \Cref{subsec:functionalSpecs}, the \texttt{Tcp\_Change\_State} procedure is the one called to implement all possible state transitions. Thus, it was selected as the procedure where we introduced a SPARK precondition that captures all the valid by the protocol's functional specification state transitions. Continuing our scenario of execution, when the procedure \texttt{Tcp\_Change\_State} is called at line \ref{bugProg:tcpChangeState}, its pre-condition warns that the states \CLOSEWAIT{} $\rightarrow$ \FINWAITONE{} and \CLOSED{} $\rightarrow$ \FINWAITONE{} are not allowed. Looking at the TCP automaton in \Cref{fig:TCPAutomaton}, we can confirm that the warning is correct since no such transitions exist on the automaton, and thus, both the C code and the SPARK translated code had this bug of allowing incorrect state transitions. The bug was also encoded on the original SPARK post-condition introduced to the \texttt{Tcp\_Wait\_For\_Events} procedure. This contract was hand-written and imprecise as, initially, not enough information was known about the possible state transitions that could be triggered by the C code. After gaining confidence about such transitions with the use of the \emph{KLEE} symbolic execution engine, as described in \Cref{subsubsect:KLEE}, the contracts that captured the bug were introduced to the \texttt{Tcp\_Change\_State}. The incorrect contracts introduced initially to the \texttt{Tcp\_Wait\_For\_Events} procedure were also corrected. The affected code was then fixed to remove the erroneous transitions, both from the C and SPARK implementations.  Finally, the SPARK prove was executed again to confirm that no more incorrect-transition warnings exist.

\section{Conclusion}

The work is building upon the previous white paper of hardening software libraries using the SPARK technology, \cite{AdaCore_Hardening_Soft_Libr_EMBENCH}, where we demonstrated how to achieve AoRTE when translating C code to Ada and SPARK. In this deliverable, we go a step further than AoRTE to illustrate how the SPARK technology can be used to prove the conformance of a software library's implementation to the library's predefined functional specifications.

Acknowledging the wide use of the TCP communication protocol in cyber-physical systems, we selected a professional-grade embedded TCP/IP library, namely the \emph{CycloneTCP} TCP, and we focused on the hardening of its TCP protocol implementation. We aimed to harden the TCP library in the areas that its original authors designated as the most vulnerable or crucial to conform to their functional specifications. Thus, the work was focused on hardening the user task's API, which allows the user to control the state of a TCP connection. More specifically, the work achieved:

\begin{itemize}
\item \textbf{Hardening the user's API} -- Using SPARK contracts, we enforced the correct usage of the \emph{CycloneTCP} TCP library's API by the user. More specifically, we enforced the library's user calls to the API functions to be in an order only allowed by the protocol's functional specification. We also ensure that the return codes of function-calls are checked for errors. Thus, in case of an error code is being returned, the application will not result in behaviour that is unspecified by the protocol specification.
\item \textbf{Conformance to the protocol's functional specification} -- We verified that the transitions allowed by the current implementation of the user-task related functions respect the state machine of \Cref{fig:TCPAutomaton}, which defines the permitted transitions between the different TCP states. We also ensured that the user's related functions are always updating a socket's state within the protocol's functional specifications.
\end{itemize}

Another significant contribution of this work is using a hybrid-approach involving the SPARK technology and a symbolic execution engine, KLEE, in our case. This hybrid approach allows proving the conformance of functional specifications related to parts of the code that are not fully translated to SPARK, but that can affect the behavior of the SPARK translated code. More specifically, we used KLEE to gain confidence that TCP state transitions enabled by code written in C respect the state transitions allowed by the TCP's functional specification. Then we introduced SPARK contracts related to those transitions that allowed the SPARK code to be proved.

Any critical application that operates outside of its functional specifications is susceptible to safety and security vulnerabilities. Thus, the approach taken in this work can be used as a guideline on how to enhance the security of software libraries written in C by using the SPARK technology to prove the conformance of the library's implementation to its functional specifications. This is perfectly aligned with the need to conform to the emerging aviation standards related to aviation cyber-security airworthiness certification, such as the ED-202A, ED-203A, DO-326A, and the DO-356A standards.

\section*{Acknowledgments}
This research is part of the ``High-Integrity Complex Large Software and Electronic Systems'' (HICLASS) project that is supported by the Aerospace Technology Institute (ATI) Programme, a joint Government and industry investment to maintain and grow the UK’s competitive position in civil aero- space design and manufacture, under grant agreement No. 113213. The programme, delivered through a partnership between the ATI, Department for Business, Energy \& Industrial Strategy (BEIS), and Innovate UK, addresses technology, capability, and supply chain challenges.

\medskip

\printbibliography

\end{document}